\documentstyle[11pt,moriond,epsfig]{article}

\def\Journal#1#2#3#4{{#1} {\bf #2}, #3 (#4)}

\def\NIMA{{\em Nucl. Instrum. Methods} A}
\def\NPB{{\em Nucl. Phys.} B}

\def\PRL{\em Phys. Rev. Lett.}

\def\be{\begin{equation}}
\def\ee{\end{equation}}
\def\bea{\begin{eqnarray}}
\def\eea{\end{eqnarray}}

\def\GeV{\mbox{~GeV}}

\def\GeVcc{\mbox{~GeV}/c^2}
\def\MeVcc{\mbox{~MeV}/c^2}

\def\BtoRG{B\to \rho\gamma}
\def\BtoRPG{B^+\to \rho^+\gamma}
\def\BtoRZG{B^0\to \rho^0\gamma}
\def\BtoROG{B\to (\rho,\omega)\gamma}
\def\BtoOG{B^0\to \omega\gamma}
\def\BtoKG{B\to K^*\gamma}
\def\BtoKZG{B^0\to K^{*0}\gamma}
\def\BtoKPG{B^+\to K^{*+}\gamma}

\def\Mbc{M_{\rm bc}}
\def\DeltaE{\Delta{E}}
\def\Ebeam{E^*_{\rm beam}{}}

\def\piZ{{\pi^0}}
\def\piP{{\pi^+}}
\def\piM{{\pi^-}}
\def\rhoP{{\rho^+}}
\def\rhoZ{{\rho^0}}
\def\KP{K^+}

\def\Kstar{{K^*}}
\def\KstarZ{{K^{*0}}}
\def\KstarP{{K^{*+}}}
\def\omegaG{{\omega\gamma}}

\def\rhoPG{{\rhoP\gamma}}
\def\rhoZG{{\rhoZ\gamma}}

\def\KstarG{{\Kstar\gamma}}

\def\Xs{X_s}

\def\MKpi{M_{K\pi}}
\def\thetahel{\theta_{\rm hel}}
\def\coshel{\cos\thetahel}
\def\Br{{\cal B}}
\def\tauBratio{{\tau_{B^+}\over\tau_{B^0}}}
\def\Lzero{{\cal L}_0}
\def\Lmax{{\cal L}_{\rm max}}

\def\PM#1#2{\,^{+#1}_{-#2}{}}
\def\EM#1{\times10^{-#1}}

\def\NRPG{15.5\PM{7.1}{6.3}\pm{1.5}}
\def\NRZG{3.6\PM{3.6}{2.8}\PM{0.7}{0.9}}
\def\NOMG{8.9\PM{4.8}{4.0}\pm{1.2}}
\def\effRPG{(5.6\pm0.4)\%}
\def\effRZG{(5.0\pm0.3)\%}
\def\effOMG{(4.7\pm0.5)\%}
\def\sROG{3.5}
\def\sRPG{2.5}
\def\sRZG{1.2}
\def\sOMG{2.3}

\def\BrROG{(1.8\PM{0.6}{0.5}\pm{0.1})\EM6}
\def\BrRPG{(1.8\PM{0.8}{0.7}\pm{0.1})\EM6}
\def\BrRZG{(0.5\PM{0.5}{0.4}\pm{0.2})\EM6}
\def\BrOMG{(1.3\PM{0.7}{0.6}\pm{0.2})\EM6}

\newcommand{\FullBR}{3.59\pm0.32\PM{0.30}{0.31}\PM{0.11}{0.07}}
\newcommand{\FullFM}{2.289\pm0.026\pm0.034}
\newcommand{\FullSM}{0.0311\pm0.0073\pm0.0063}

\def\etal{\textit{et al.}}
\def\Journal#1#2#3#4{{#1} {\bf #2}, #3 (#4)} % {journal}{vol}{page}{year}
\def\NIMA{Nucl. Instrum. Meth. A}
\def\NPB{Nucl. Phys. B}

\def\PRL{Phys. Rev. Lett.}

\def\EPJC{Eur. Phys. J. C}

\begin{document}
\vspace*{4cm}
\title{Measurements of Electroweak Penguin Processes $b \to d,s
  \gamma$ and $b \to s\ell^+\ell^-$ at Belle}

\author{M.Iwasaki \\
         ( for the Belle collaboration )} 

\address{Dept. of Physics, University of Tokyo,
  7-3-1 Hongo, Bunkyo-ku, Tokyo 113-0011 JAPAN}

\maketitle\abstracts{
We report the measurements of electroweak penguin processes 
$b \to d,s \gamma$ and $b \to s\ell^+\ell^-$
at the Belle experiment. The analyses are based on 152 million 
$B\overline{B}$ events collected at the $\Upsilon(4S)$ resonance with
the Belle detector at the KEKB $e^+e^-$ asymmetric-energy collider.
The $b \to d \gamma$ transition is studied through 
exclusive decays: $B^0 \to \rho^{0} \gamma$, $B^0 \to \omega \gamma$
and $B^+ \to \rho^+ \gamma$. We apply fully inclusive and 
semi-inclusive reconstruction to $b \to s \gamma$ and 
$b \to s\ell^+\ell^-$ processes, respectively.
The results on the exclusive 
$b \to d\gamma$ and semi-inclusive $b \to s\ell^+\ell^-$ 
measurements are preliminary.
}
\section{Introduction}
In the Standard Model (SM), flavor-changing neutral current (FCNC)
process is forbidden at tree level. 
Such a process can occur at higher order via electroweak penguin 
and box diagrams.
The FCNC process is therefore sensitive to the effects of 
non-SM physics that may enter these loops.

In this report, 
we show the measurements of FCNC processes 
$b \to d,s \gamma$ and $b \to s\ell^+\ell^-$.
The $b \to d \gamma$ process is studied through 
exclusive decays: $B^0 \to \rho^{0} \gamma$, $B^0 \to \omega \gamma$
and $B^+ \to \rho^+ \gamma$. We apply fully inclusive and 
semi-inclusive reconstruction to $b \to s \gamma$ and 
$b \to s\ell^+\ell^-$ processes, respectively.
These measurements are based on a sample of $152 \times
10^{6}$~$\Upsilon(4S) \to B\overline{B}$ decays
collected with the Belle detector~\cite{Belle} 
at the KEKB~\cite{KEKB} $e^+e^-$ asymmetric-energy collider.

\section{Exclusive $b \to d \gamma$}
The $b \to d \gamma$ process is suppressed with respect to
$b \to s \gamma$ by the Cabibbo-Kobayashi-Maskawa (CKM) factor
$|V_{td}/V_{ts}|^2 \sim 0.04$ with a large uncertainty due to lack of
precise knowledge on $V_{td}$.

In this analysis, we use the exclusive reconstruction
of the decays, $\BtoRPG$, $\BtoRZG$, and $\BtoOG$.\footnote{
Throughout the article charge-conjugate 
states are implicitly included.}
We also reconstruct $\BtoKPG$ and $\BtoKZG$ for control samples.  
The following decay chains are used to reconstruct the intermediate states:
$\rhoP\to\piP\piZ$, $\rhoZ\to\piP\piM$, $\omega\to\piP\piM\piZ$,
$\KstarP\to\KP\piZ$, $\KstarZ\to\KP\piM$, and $\piZ\to\gamma\gamma$.
In each event, we select the photon with the largest energy 
in the range $1.8\GeV<E_{\gamma}<3.4\GeV$, 
in the $e^{+}e^{-}$ center-of-mass frame (CM).
To suppress backgrounds from $\piZ\to\gamma\gamma$ and 
$\eta\to\gamma\gamma$ decays,
we veto the event if the reconstructed mass of the primary photon and
any other photon of 30 (200) MeV or more is within $\pm18$ $(32)\MeVcc$
of the $\piZ$ ($\eta$) mass. 
$B$ candidates are formed by combining a $\rho$ or $\omega$ 
candidate and the primary photon using two variables: the beam-energy
constrained mass $\Mbc = \sqrt{ (\Ebeam/c^2)^2 - |p_{B}^*/c|^{2}}$ and
the energy difference $\Delta E = E^*_{B} - \Ebeam$, where $p^*_{B}$ and
$E^*_B$ are the measured CM momentum and energy, respectively, of the
$B$ candidate, and $\Ebeam$ is the CM beam energy.  The photon energy is
replaced by $\Ebeam - E_{\rho/\omega}^*$ if the momentum $p^*_{B}$ is
calculated.  We define the signal region as $-0.1\GeV<\DeltaE<0.08\GeV$
and $5.273\GeVcc<\Mbc<5.285\GeVcc$. 

There are two major background sources from $B$ decays: 
$\BtoKG$, and $B\to\rho/\omega\piZ$.
To suppress $\BtoKG$, we calculate
$\MKpi$, where a kaon mass is assigned to one of the pion candidate, and
reject the candidate if $\MKpi<0.96$ ($0.92$) $\GeVcc$ for the $\rhoZG$
($\rhoPG$) mode.  
To reject $B\to\rho/\omega\piZ$, we apply a helicity angle cut
%cosine of the helicity angle $\thetahel$, and reject the candidate if
$|\coshel|>0.8$ ($0.6$) for $\rhoZG$ and $\omegaG$ ($\rhoPG$) modes.
Here, $\thetahel$ is the angle between $\piP$ and $B$
momentum vectors in the $\rho$ rest frame, or between the
normal to the $\omega$ decay plane and the $B$ momentum vector in the
$\omega$ rest frame.

The background 
from continuum $e^+e^-\to q\bar{q}$ ($q=u,d,s,c$) events 
is rejected using the event topology information.
We use a Fisher discriminant\cite{bib:fisher} 
constructed from 16 modified Fox-Wolfram
moments~\cite{bib:fox-wolfram} 
and the scalar sum of the transverse momentum~\cite{bib:belle-pi0pi0}.  
We also use the decay vertex of the candidate $B$ meson as well as the
origin of the remaining tracks in the event.
The difference between these two vertices along the
$z$-axis discriminates continuum events that has a common decay vertex 
and signal events whose decay vertices are displaced in the laboratory frame.

To obtain the signal yield, we perform an unbinned maximum likelihood
fit to $\Mbc$ and $\DeltaE$ to the events in $|\DeltaE|<0.3\GeV$ and
$\Mbc>5.2\GeVcc$.
We perform a simultaneous fit to three signal modes 
($B \to (\rho,\omega)\gamma$)
plus two $\BtoKG$ modes, 
assuming the isospin relations $\Br(\BtoROG) \equiv
\Br(\BtoRPG) = 2\tauBratio\Br(\BtoRZG) = 2\tauBratio\Br(\BtoOG)$ and
$\Br(\BtoKPG) = \tauBratio\Br(\BtoKZG)$, where we use $\tauBratio =
1.083\pm0.017$ \cite{bib:pdg2002}.  Here, floated parameters are the
branching fractions for $\BtoROG$ and $\BtoKG$, five background
normalizations and five background $\DeltaE$ slopes.  
%We assume the efficiency systematic errors are fully correlated.

\begin{figure}%[bpht]
\begin{center}
\psfig{figure=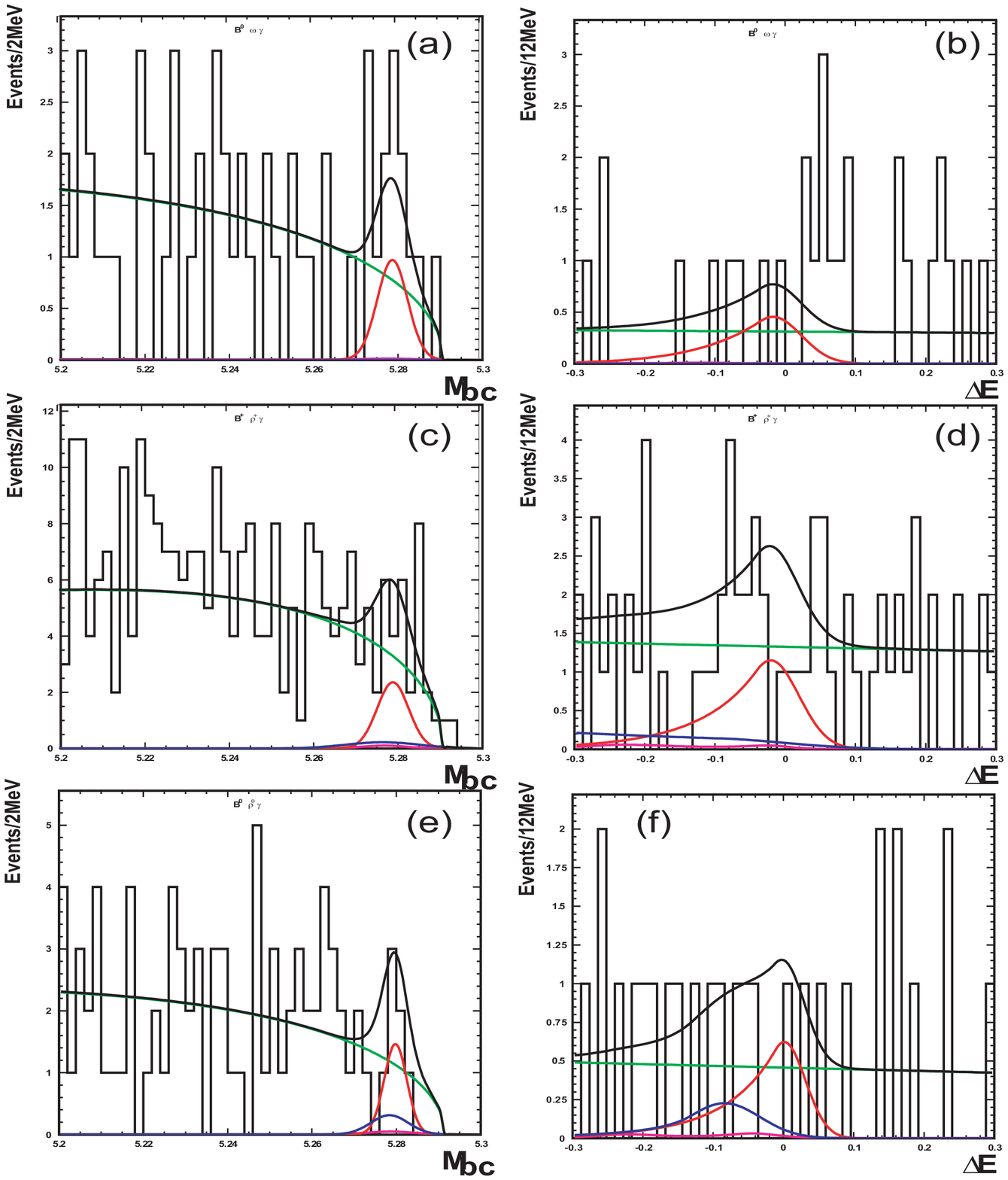,height=10cm}
\end{center}
\caption{Projections of the simultaneous fit results to $\Mbc$ and
  $\DeltaE$ of the individual modes to the signal region, 
  for 
  $\BtoOG$  (a and b),  
  $\BtoRPG$ (c and d), and 
  $\BtoRZG$ (e and f). 
  Lines
  represent the total fit result (black), signal (red), $\KstarG$
  (blue) and continuum (green) components.
  $B\to(\rho/\omega)\piZ$ components are invisibly small.}
\label{fig:bdgam}
\end{figure}

Preliminary results of the simultaneous fit 
are shown in Fig.~\ref{fig:bdgam} and
given in Table~\ref{table:bdgam}.  We find the simultaneous fit gives a
significance of 3.5, where significance is defined as
$\sqrt{-2\ln(\Lzero/\Lmax)}$, $\Lmax$ is the maximum likelihood in the
$\Mbc$ fit, and $\Lzero$ is the likelihood of the best fit here the
signal yield is constrained to be zero.

\begin{table}[b]
\begin{center}
\begin{tabular}{lcccc}
\hline \hline
Mode & efficiency & signal yield & significance 
     & branching fraction \\
{}   & ($\pm$syst) & ($\pm$stat$\pm$syst) & & ($\pm$stat$\pm$syst) \\
\hline
$\BtoRPG$ &$\effRPG$ &$\NRPG$ &$\sRPG$ &$\BrRPG$ \\
$\BtoRZG$ &$\effRZG$ &$\NRZG$ &$\sRZG$ &$\BrRZG$ \\
$\BtoOG$  &$\effOMG$ &$\NOMG$ &$\sOMG$ &$\BrOMG$ \\
$\BtoROG$ &---       &---     &$\sROG$ &$\BrROG$ \\
\hline \hline
\end{tabular}
\end{center}
\caption{ Results of the efficiency, signal yield, significance, and 
  branching fraction from simultaneous and individual fits.
  These numbers are preliminary.}
\label{table:bdgam}
\end{table}

\section{Inclusive $b \to s \gamma$}
We recently measured the $b \to s \gamma$ branching fraction
using a fully-inclusive approach.
The detailed descriptions are written in Ref.~\cite{bsgam}
In this analysis, to extract the signal $b \to s \gamma$ spectrum, 
we collect all high-energy photons, 
vetoing those originating from $\pi^0$ and $\eta$ 
decays to two photons. The contribution 
from continuum events 
is subtracted using the off-resonance sample. The remaining 
backgrounds from $B\bar{B}$ events are subtracted using 
Monte-Carlo (MC) distributions scaled by data control samples.
After subtracting the backgrounds, we correct the
photon spectrum for the signal selection efficiency function
obtained from signal MC, applying the correction determined 
by data control samples.
%The average signal selection efficiency is $23$\%.

\begin{figure}
\centerline{
\epsfxsize 3.0 truein \epsfbox{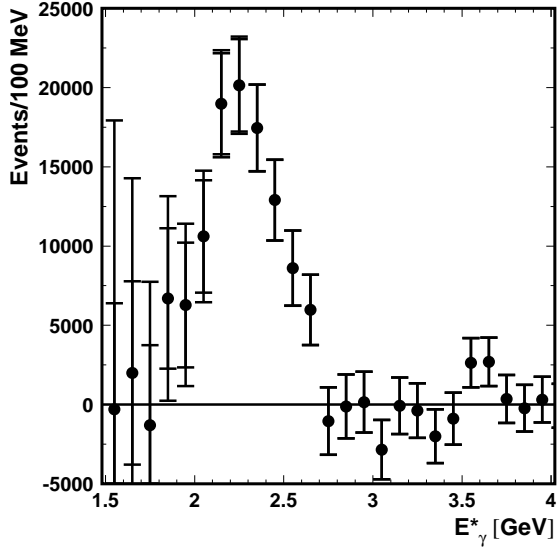}
}   
\caption{Efficiency-corrected photon energy spectrum.
         The two error bars show the statistical and total errors.}
\label{fig:figure2}
\end{figure}

The efficiency-corrected spectrum is shown 
as a function of CM photon energy
in Figure~\ref{fig:figure2}. 
The two error bars for each point show the statistical
and the total error, including the systematic error which is correlated among the points.
As expected, the spectrum above the $3\,\GeV$ endpoint 
for decays of $B$ mesons from the $\Upsilon(4S)$ 
is consistent with zero.
Integrating this spectrum from $1.8$ to $2.8\,\GeV$, we obtain a
partial branching fraction of 
${\Br(b \to s\gamma)} = \left(\FullBR\right)\times 10^{-4}$,
where the errors are statistical, systematic and theoretical, respectively. 
This result is in good agreement with the latest theoretical 
calculations~\cite{Hurth:2003dk,Gambino:2001ew}.
We have also measured the moments of the distribution and obtain
$\left< E_\gamma \right>  =   \FullFM\,\GeV$ and 
$\left<E_\gamma^2\right>-\left<E_\gamma\right>^2 = \FullSM \,\GeV^2$ 
for $E^\ast_\gamma>1.8\,\GeV$,
where the errors are statistical and systematic. 

\section{Semi-inclusive $b \to s \ell^+ \ell^-$}
We study the inclusive $B \to X_{s} \ell^+ \ell^-$ process,
where $\ell$ is an electron or a muon
and $\Xs$ is a hadronic system containing an $s$-quark.
The $s$-quark hadronic system is reconstructed with
a semi-inclusive reconstruction approach:  
reconstruct with one $K^{\pm}$ or $K^{0}_{s}$ 
and up to four pions (at most one pion can be neutral).
Compared to a fully inclusive approach, this method has the advantage
of having the strong kinematical discrimination by using the 
beam-energy constrained mass 
$\Mbc$ and the energy difference $\DeltaE$.

In addition to the discrimination, further background suppression
to reduce the large combinatorial backgrounds is necessary. 
The main contribution to the combinatorial background is
from semileptonic decays in $B\bar{B}$ events.
In these events, $B \to X_{s} \ell^+ \ell^-$ candidates 
are reconstructed with
the decay products from both $B\bar{B}$ mesons .
Such a background event has a significant amount of missing energy
due to the neutrinos from the semileptonic decays.
Another contribution to the combinatorial background is from
continuum events, which are effectively suppressed 
with event-shape variables.

There are two background sources that can peak in $\Mbc$ and
$\DeltaE$.
The first is  from
$B \to J/\psi X$ and $B \to \psi(2S) X$ decays with
$J/\psi (\psi(2S)) \to \ell^{+} \ell^{-}$.
This peaking background is efficiently removed with cuts on the 
dilepton mass $m(\ell^{+} \ell^{-})$. The resulting veto sample provides
a large control sample of decays with a signature identical to that
of the signal.
The second is from $B \to K^{\pm} (K^0_s) n \pi$~$(n > 1)$ 
decays with misidentification of two charged pions as leptons.
We estimate these peaking background contaminations, then
subtract them from the signal yield.

\begin{figure}[thp]
\begin{center}
\psfig{figure=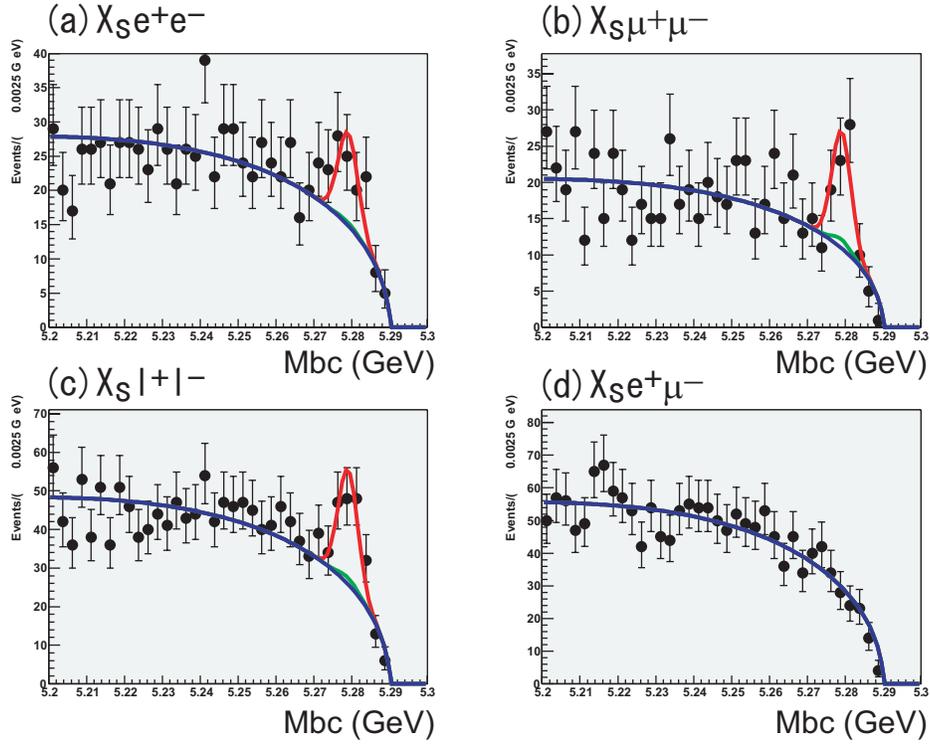,height=10cm}
\end{center}
\caption{Distributions of $M_{bc}$ for selected
(a) $B \to X_s e^+e^-$, 
(b) $B \to X_s \mu^+\mu^-$, 
(c) $B \to X_s \ell^+\ell^-$,  ($\ell = e, \mu$), 
and 
(d) $B \to X_s e^\pm \mu^\mp$ candidates.
The red lines represent the result of the fits,
and the green and blue lines represent 
the peaking and combinatorial background 
components under the signal peaks, respectively.}
\label{fig:xsll1}
\end{figure}

We perform an extended, unbinned maximum likelihood fit
to the $M_{bc}$ distribution 
in the region $M_{bc} > 5.2$$\GeVcc$ to extract the signal yield as well as
the shape and yield of the combinatorial background.
We fit the $M_{bc}$ distributions for the selected
$B \to X_s e^+ e^-$ and $B \to X_s \mu^+ \mu^-$ candidates separately,
and  $B \to X_{s} \ell^+ \ell^-$ ($\ell = e, \mu$).
The obtained results are shown in Figures~\ref{fig:xsll1}(a)-(c).
Figure~\ref{fig:xsll1}(d) shows the distribution for 
$B \to X_s e^\pm \mu^\mp$ candidates. 
There is no peaking component as expected.
Using a sample of $152 \times 10^{6}$ $B\overline{B}$ decays, 
we observe a signal of $72.3 \pm 13.8(stat) \pm 4.6(syst)$ events.

\begin{figure}[thp]
\begin{center}
\psfig{figure=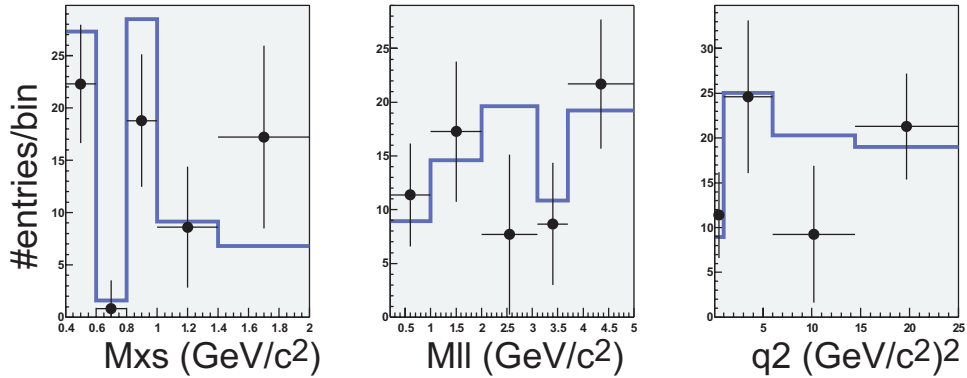,height=5cm}
\end{center}
\caption{Distributions of number of signal events as a function
 of (a) hadronic mass $M_{xs}$ (b) lepton pair mass $M_{ll}$, 
and (c) $q^2 \equiv M_{ll}^2$ for electron and muon channels combined
 for data (points) and Monte Carlo signal (histogram).
 The vertical error bars represent statistical errors only.}
\label{fig:xsll2}
\end{figure}

Figures~\ref{fig:xsll2}(a)-(c) show the distributions of
hadronic mass $M_{xs}$, lepton pair mass $M_{ll}$, 
and $q^2 \equiv M_{ll}^2$ for electron and muon channels combined,
obtained by performing the nominal likelihood fit
in separate $M_{xs}$, $M_{ll}$, and $q^2$ regions.

The preliminary results of the measured branching fractions
for $M(\ell^{+} \ell^{-}) > 0.2$$\GeVcc$ are
\begin{eqnarray*}
 {\cal B}(B \to X_s e^+ e^-)   
 & = & \left(4.45 \pm 1.32 ^{+0.84}_{-0.79}\right)
 \times 10^{-6}, \\
 {\cal B}(B \to X_s \mu^+ \mu^-) 
 & = & \left(4.30 \pm 1.06 ^{+0.74}_{-0.70}\right)
 \times 10^{-6}, \mbox{and} \\
 {\cal B}(B \to X_s \ell^+ \ell^-)   
 & = & \left(4.39 \pm 0.84 ^{+0.78}_{-0.73}\right)
 \times 10^{-6},
\end{eqnarray*}
where the first error is statistical and the second error is systematic.
The combined $B \to X_s \ell^+ \ell^-$ branching fraction is 
the weighted average of
the branching fractions for the electron and muon channels,
where we assume the individual branching fractions to be equal 
for $M(\ell^{+} \ell^{-}) > 0.2$$\GeVcc$.

\section{Summary}
We have studied the electroweak penguin processes 
$b \to d,s \gamma$ and $b \to s\ell^+\ell^-$
based on a sample of $152 \times 10^{6}$
$\Upsilon(4S) \to B\overline{B}$ decays
at the Belle experiment.
The $b \to d \gamma$ transition is studied through 
exclusive decays: $B^0 \to \rho^{0} \gamma$, $B^0 \to \omega \gamma$
and $B^+ \to \rho^+ \gamma$. We apply fully inclusive and 
semi-inclusive reconstruction to $b \to s \gamma$ and 
$b \to s\ell^+\ell^-$ processes, respectively.

We present the first evidence for the $b \to d\gamma$ process
using a simultaneous fit to the $\BtoRG$ and $\BtoOG$ modes.  
The preliminary result is consistent with 
the SM predictions~\cite{bib:ali-parkhomenko,bib:bosch-buchalla}.
We measure the branching fraction and photon energy spectrum of
inclusive $b \to s\gamma$ process in the
energy range $1.8\,\GeV\le E^\ast_\gamma\le2.8\,\GeV$.
For the first time $95$\% or more of the spectrum
is measured, allowing the theoretical uncertainties to be
reduced to a very low level. 
We measure the branching fraction for the $B \to X_s \ell^+\ell^-$, 
where $\ell = e$ or $\mu$ 
and $\Xs$ hadronic system is semi-inclusively 
reconstructed.
The preliminary measurement is consistent with the recent prediction
by Ali {\em et al.}~\cite{Ali02}.

\section*{Acknowledgments}
We wish to thank the KEKB accelerator group for excellent operation of
the KEKB accelerator.

\end{document}